\documentclass[twoside,twocolumn,9pt]{article}
\usepackage{extsizes}
\usepackage[super,sort&compress,comma]{natbib}
\usepackage[version=3]{mhchem}
\usepackage[left=1.5cm, right=1.5cm, top=1.785cm, bottom=2.0cm]{geometry}
\usepackage{balance}
\usepackage{mathptmx}
\usepackage{sectsty}
\usepackage{graphicx}
\usepackage{lastpage}
\usepackage[format=plain,justification=justified,singlelinecheck=false,font={stretch=1.125,small,sf},labelfont=bf,labelsep=space]{caption}
\usepackage{float}
\usepackage{fancyhdr}
\usepackage{fnpos}
\usepackage[english]{babel}
\addto{\captionsenglish}{%
  
}
\usepackage{array}
\usepackage{droidsans}
\usepackage{charter}
\usepackage[T1]{fontenc}
\usepackage[usenames,dvipsnames]{xcolor}
\usepackage{setspace}
\usepackage[compact]{titlesec}
\usepackage[hidelinks]{hyperref}
\usepackage{epstopdf}

\usepackage{amsmath}
\usepackage{amssymb}
\usepackage{multirow}
\IfFileExists{orcidlink.sty}{\usepackage{orcidlink}}{\newcommand{\orcidlink}[1]{}}

\definecolor{cream}{RGB}{222,217,201}

\usepackage{xcolor}
\definecolor{revcol}{rgb}{0.00,0.00,0.00}  
\newcommand{\rev}[1]{\textcolor{revcol}{#1}}

\begin{document}

\pagestyle{fancy}
\thispagestyle{plain}
\fancypagestyle{plain}{
\renewcommand{\headrulewidth}{0pt}
}

\makeFNbottom
\makeatletter
\renewcommand\LARGE{\@setfontsize\LARGE{15pt}{17}}
\renewcommand\Large{\@setfontsize\Large{12pt}{14}}
\renewcommand\large{\@setfontsize\large{10pt}{12}}
\renewcommand\footnotesize{\@setfontsize\footnotesize{7pt}{10}}
\makeatother

\renewcommand{\thefootnote}{\fnsymbol{footnote}}
\renewcommand\footnoterule{\vspace*{1pt}%
\color{cream}\hrule width 3.5in height 0.4pt \color{black}\vspace*{5pt}}
\setcounter{secnumdepth}{5}

\makeatletter
\renewcommand\@biblabel[1]{#1}
\renewcommand\@makefntext[1]%
{\noindent\makebox[0pt][r]{\@thefnmark\,}#1}
\makeatother
\renewcommand{\figurename}{\small{Fig.}~}
\sectionfont{\sffamily\Large}
\subsectionfont{\normalsize}
\subsubsectionfont{\bf}
\setstretch{1.125}
\setlength{\skip\footins}{0.8cm}
\setlength{\footnotesep}{0.25cm}
\setlength{\jot}{10pt}
\titlespacing*{\section}{0pt}{4pt}{4pt}
\titlespacing*{\subsection}{0pt}{15pt}{1pt}

\fancyfoot{}
\fancyfoot[LO,RE]{\vspace{-7.1pt}\includegraphics[height=9pt]{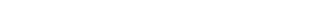}}
\fancyfoot[CO]{\vspace{-7.1pt}\hspace{13.2cm}\includegraphics{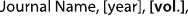}}
\fancyfoot[CE]{\vspace{-7.2pt}\hspace{-14.2cm}\includegraphics{head_foot/RF}}
\fancyfoot[RO]{\footnotesize{\sffamily{1--\pageref{LastPage} ~\textbar  \hspace{2pt}\thepage}}}
\fancyfoot[LE]{\footnotesize{\sffamily{\thepage~\textbar\hspace{3.45cm} 1--\pageref{LastPage}}}}
\fancyhead{}
\renewcommand{\headrulewidth}{0pt}
\renewcommand{\footrulewidth}{0pt}
\setlength{\arrayrulewidth}{1pt}
\setlength{\columnsep}{6.5mm}
\setlength\bibsep{1pt}

\makeatletter
\newlength{\figrulesep}
\setlength{\figrulesep}{0.5\textfloatsep}

\newcommand{\topfigrule}{\vspace*{-1pt}%
\noindent{\color{cream}\rule[-\figrulesep]{\columnwidth}{1.5pt}} }

\newcommand{\botfigrule}{\vspace*{-2pt}%
\noindent{\color{cream}\rule[\figrulesep]{\columnwidth}{1.5pt}} }

\newcommand{\dblfigrule}{\vspace*{-1pt}%
\noindent{\color{cream}\rule[-\figrulesep]{\textwidth}{1.5pt}} }

\makeatother

\setcounter{topnumber}{3}
\setcounter{bottomnumber}{2}
\setcounter{totalnumber}{4}
\setcounter{dbltopnumber}{3}
\renewcommand{\topfraction}{0.90}
\renewcommand{\bottomfraction}{0.70}
\renewcommand{\dbltopfraction}{0.90}
\renewcommand{\textfraction}{0.07}
\renewcommand{\floatpagefraction}{0.75}
\renewcommand{\dblfloatpagefraction}{0.75}

\setlength{\textfloatsep}{10pt plus 2pt minus 2pt}
\setlength{\dbltextfloatsep}{10pt plus 2pt minus 2pt}
\setlength{\floatsep}{8pt plus 2pt minus 2pt}
\setlength{\dblfloatsep}{8pt plus 2pt minus 2pt}
\setlength{\intextsep}{8pt plus 2pt minus 2pt}
\captionsetup{aboveskip=5pt,belowskip=0pt}

\newcommand{\panel}[1]{\\[1pt]{\small\sffamily (#1)}}

\twocolumn[
  \begin{@twocolumnfalse}
{\includegraphics[height=30pt]{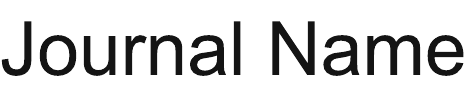}\hfill\raisebox{0pt}[0pt][0pt]{\includegraphics[height=55pt]{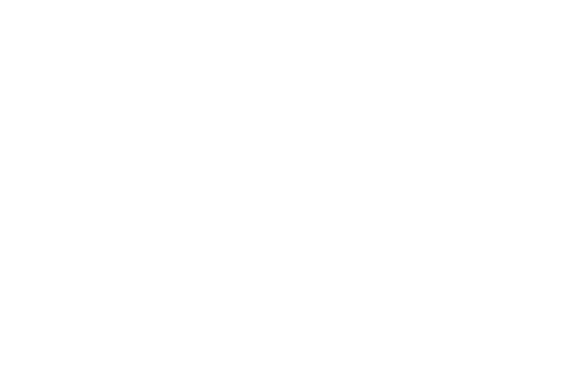}}\\[1ex]
\includegraphics[width=18.5cm]{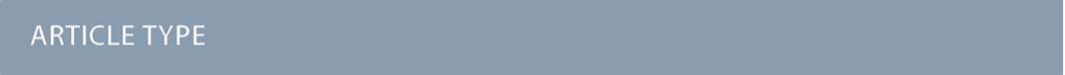}}\par
\vspace{1em}
\sffamily
\begin{tabular}{m{4.5cm} p{13.5cm} }

\includegraphics{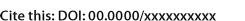} & \noindent\LARGE{\textbf{\rev{Polarization-multiplexed spatial filtering in a symmetric TiO$_2$ metasurface with an anisotropic van der Waals gap fill}$^{\dag}$}} \\

\vspace{0.3cm} & \vspace{0.3cm} \\

 & \noindent\large{Shoumik Debnath\,\orcidlink{0009-0001-9500-5319}\textit{$^{a}$} and Sudipta Saha\,\orcidlink{0009-0003-9110-8059}$^{\ast}$\textit{$^{a,b}$}} \\

\includegraphics{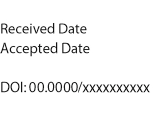} & \noindent\normalsize{\rev{Finite-difference time-domain simulations of a symmetric TiO$_2$ nanobar-pair metasurface with a 60~nm $\alpha$-MoO$_3$ gap fill give two Fano resonances that are selected by the input polarization angle, at 885.5~nm ($Q=50.5$) and 893.8~nm ($Q=44.8$).} \rev{Both channels act as spatial high-pass filters and differ in their cutoff spatial frequency, 0.44 and 1.17~$\mu$m$^{-1}$, with the broader cutoff belonging to the lower-$Q$ resonance.} \rev{The unit cell retains a mirror plane normal to the swept direction, so the momentum-space transfer function is even in $k_x$; over the accessible momentum range the broader channel follows $|H|\propto|k_x|$, which is the kernel used for edge enhancement.} Both operations are applied to a USAF~1951 resolution chart in a simulated 4$f$ arrangement. \rev{The simulations use measured anisotropic optical constants for $\alpha$-MoO$_3$ with the crystal axes assigned as they lie in an exfoliated flake.} \rev{A control series with isotropic and uniaxial gap fillings indicates that the shape anisotropy of the bar pair sets the scale of the channel separation, that the anisotropy of the crystal acts mainly through its out-of-plane index contrast, and that the in-plane birefringence contributes a small part of it.}} \\
\end{tabular}

 \end{@twocolumnfalse} \vspace{0.6cm}

  ]

\renewcommand*\rmdefault{bch}\normalfont\upshape
\rmfamily
\section*{}
\vspace{-1cm}

\footnotetext{\textit{$^{a}$~Department of Electrical and Electronic Engineering, Bangladesh University of Engineering and Technology (BUET), Dhaka 1205, Bangladesh.}}
\footnotetext{\textit{$^{b}$~Accident Research Institute, Bangladesh University of Engineering and Technology (BUET), Dhaka 1000, Bangladesh. E-mail: sudiptasaha@ari.buet.ac.bd}}
\footnotetext{\dag~\rev{Supplementary Information available: simulation settings, mesh convergence, gap-filling control series, absorptive quality factor, transfer-function symmetry, polarization crosstalk, geometric tolerances, operation fits, image-processing settings and efficiency figures (Tables~S1 to S5). See DOI: 00.0000/00000000.}}

\section{Introduction}

Analog optical computing performs image-processing operations in the optical domain and removes the repeated optical-to-electrical conversion steps that a digital implementation requires.\cite{silva2014,zangeneh2021} Spatial differentiation and high-pass filtering are among the operations studied most closely, because both enhance edges and extract features used in machine vision.\cite{he2022,cui2026,yu2026,guo2018,zhu2017} Early work used plasmonic structures and photonic crystal slabs.\cite{zhu2017,guo2018} Dielectric metasurfaces have since become the more common platform, since their low absorption and high transmission allow strong modulation of the transfer function without a large throughput penalty.\cite{zhou2020,zhou2019,cordaro2019,zong2023,cotrufo2023,deng2024,swartz2024,kendall2025}

Many dielectric analog-computing platforms use Fano resonances, and in the high-$Q$ limit quasi-bound states in the continuum (quasi-BICs), to produce sharp spectral features.\cite{limonov2017,miroshnichenko2010,hsu2016,koshelev2018,kang2023} The linewidth of such a resonance is set by the coupling between a nominally dark mode and the radiation continuum. Within temporal coupled-mode theory the radiative decay rate scales as $\gamma_\mathrm{rad}\propto\rev{V}^2$, where \rev{$V$} is the coupling coefficient between the mode and the radiation channel.\cite{fan2003} Controlling the perturbation that sets this coupling therefore controls the linewidth, and through it the momentum-space response. \rev{The same principle underlies resonant sensing and analog signal processing.\cite{tittl2018,pan2021,liu2024}}

Most quasi-BIC metasurfaces break the geometric symmetry of the unit cell using shifted apertures, asymmetric resonators or truncated cells.\cite{koshelev2018} Once such a structure is fabricated, the perturbation strength is fixed. \rev{Introducing an anisotropic material into the resonator environment offers a second route, in which the optical response is modified without changing the geometry.\cite{yang2025,gomis2017,chern2023,gupta2026} Van der Waals crystals are convenient in this role because their anisotropy is a property of the bond geometry rather than of quantum confinement, so it persists from bulk single crystals down to flakes of a few tens of nanometres.\cite{caldwell2019,zheng2019} $\alpha$-MoO$_3$ has recently been characterised across the ultraviolet to near-infrared range by spectroscopic ellipsometry, which provides a measured anisotropic dielectric tensor in the band of interest here.\cite{toksumakov2026}}

This work examines a symmetric TiO$_2$ nanobar-pair metasurface with a 60~nm $\alpha$-MoO$_3$ gap fill. \rev{The two polarization channels of the cell support Fano resonances at different wavelengths and with different linewidths, and both act as spatial high-pass filters that differ in their cutoff spatial frequency.} \rev{Over the accessible momentum range the broader channel follows $|H|\propto|k_x|$, which is the regime used for edge enhancement.} Channel selection is by input polarization angle, with the same unit cell in both cases. Dual-polarization analog computing has previously been implemented with polarization-insensitive nonlocal metasurfaces,\cite{kwon2020} dispersion-engineered structures\cite{cotrufo2023} and polarization-multiplexed designs,\cite{bi2025} in which separate structural elements or dispersion engineering were used for each channel.

\begin{figure*}[t]
\centering
\begin{minipage}[b]{0.70\textwidth}
  \centering
  \includegraphics[width=\linewidth]{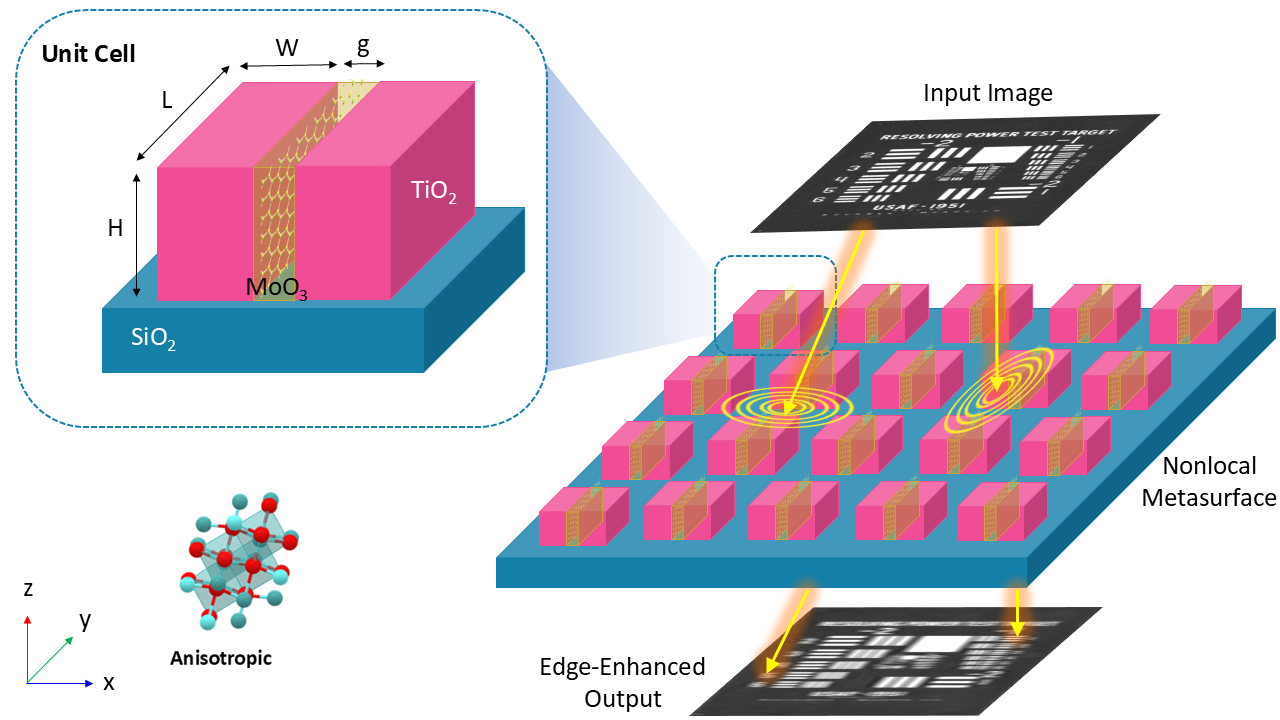}\panel{a}
\end{minipage}
\hfill
\begin{minipage}[b]{0.65\textwidth}
  \centering
  \includegraphics[width=\linewidth]{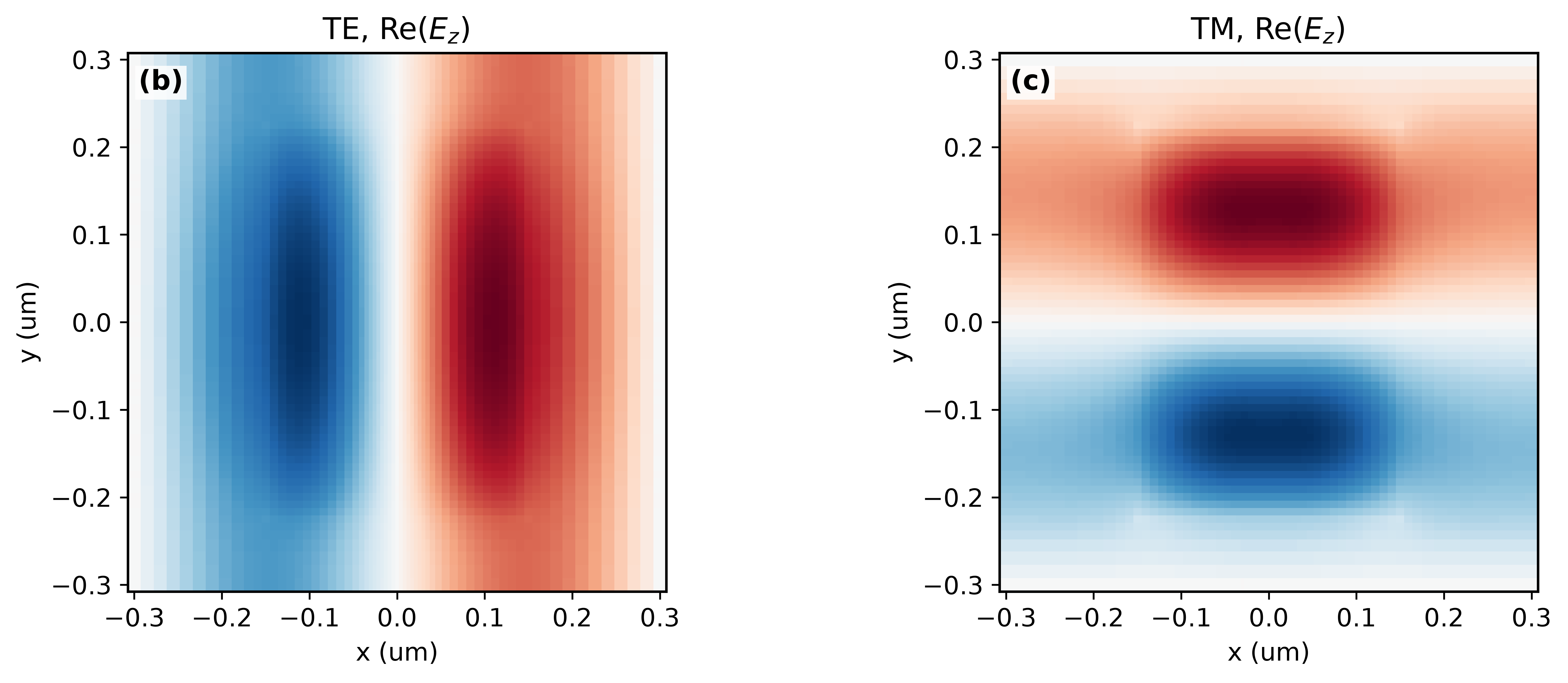}
\end{minipage}
\caption{(a) Schematic of the TiO$_2$ nanobar pair unit cell with a 60~nm $\alpha$-MoO$_3$ gap fill on a SiO$_2$ substrate (not to scale). Inset: crystal structure of $\alpha$-MoO$_3$ (orthorhombic, $Pnma$; Materials Project mp-20589\cite{horton2025accelerated,jain2013commentary}), with the three inequivalent Mo--O bond directions labelled. \rev{(b, c) Simulated $E_z$ distribution in the plane $z=H/2$ at normal incidence, for the TE and TM channels respectively.}}
\label{fig:device}
\end{figure*}

\section{Device design and simulation}

\subsection{Unit cell geometry}

The unit cell is a pair of TiO$_2$ nanobars separated by a 60~nm gap on a SiO$_2$ substrate ($n=1.46$), with the gap filled by $\alpha$-MoO$_3$ [Fig.~\ref{fig:device}(a)]. The geometric parameters are bar width $W=120$~nm, bar length $L=450$~nm, bar height $H=250$~nm, gap size $g=60$~nm and lattice period $P=600$~nm.

\rev{The bare nanobar pair has $C_{2v}$ symmetry, with mirror planes normal to $x$ and to $y$.} \rev{Because $W\neq L$, the two polarization channels are not degenerate before the gap is filled.} \rev{The gap material shifts both resonances and changes both linewidths.} \rev{Section~3.2 separates the part of that shift due to the anisotropy of the crystal from the part due to the index of the filler.}

\subsection{Material assignment and crystal orientation}

The TiO$_2$ refractive index is $n=2.35$, treated as isotropic.\cite{kischkat2012}

\rev{For $\alpha$-MoO$_3$ we use the measured anisotropic optical constants of Toksumakov \textit{et al.}\cite{toksumakov2026} in place of the calculated values of Lajaunie \textit{et al.},\cite{lajaunie2013} and we assign the crystal axes according to the orientation an exfoliated flake takes.} \rev{$\alpha$-MoO$_3$ is orthorhombic with lattice parameters 3.70, 3.96 and 13.86~\AA.} \rev{The long axis is the van der Waals stacking direction, so in a flake lying on a substrate the two short axes lie in the plane and the long axis is out of plane.\cite{toksumakov2026} The in-plane axes are accordingly assigned to $x$ and $y$ and the stacking axis to $z$.} \rev{An assignment that places the stacking axis in the plane of the flake does not correspond to an exfoliated geometry.}

\begin{figure*}[t]
\centering
\includegraphics[width=0.82\textwidth]{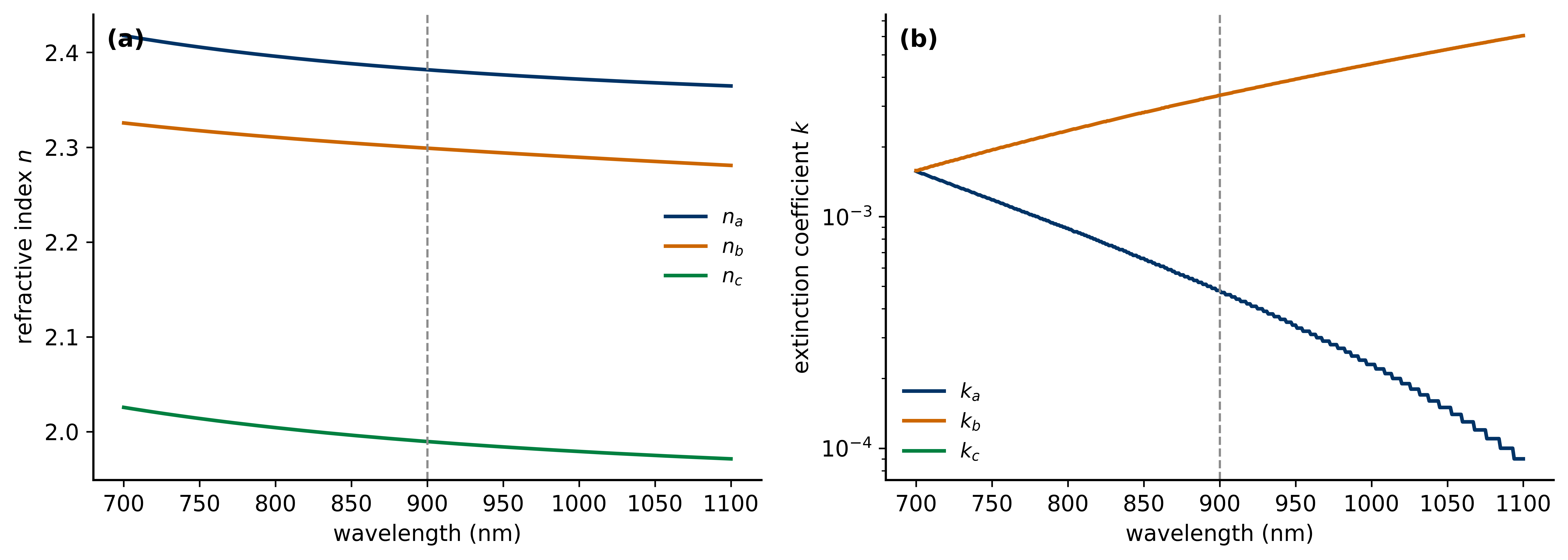}
\caption{\rev{Measured anisotropic optical constants of $\alpha$-MoO$_3$ from Toksumakov \textit{et al.}\cite{toksumakov2026} (a) Refractive index along the three crystallographic axes. (b) Extinction coefficient, logarithmic scale. The dashed line marks 900~nm, where $n_a=2.382$, $n_b=2.299$ and $n_c=1.990$. The $a$ and $b$ axes lie in the plane of an exfoliated flake and $c$ is the van der Waals stacking direction.}}
\label{fig:moo3}
\end{figure*}

\rev{Fig.~\ref{fig:moo3} shows the measured indices and extinction coefficients over the band used here.} \rev{At 900~nm the in-plane values are $n_a = 2.382$ and $n_b = 2.299$ and the out-of-plane value is $n_c = 1.990$, giving an in-plane birefringence of 0.083 and an out-of-plane contrast of 0.392.} \rev{The extinction coefficients are $k_a = 4.8\times10^{-4}$, $k_b = 3.3\times10^{-3}$ and $k_c < 1\times10^{-4}$.} The gap material is entered as a diagonal anisotropic medium with $\varepsilon_{xx}=n_a^2$, $\varepsilon_{yy}=n_b^2$ and $\varepsilon_{zz}=n_c^2$. \rev{TE denotes the channel with the electric field along $x$ and TM the channel with the field along $y$.} \rev{These labels are used as defined here: for the angular sweep of Section~2.3, which tilts the illumination in the $x$--$z$ plane, the channel labelled TE is the one whose electric field lies in the plane of incidence.}

\rev{The in-plane index of the crystal sits close to that of the TiO$_2$ bars, $2.382$ against $2.35$, while the out-of-plane index is well below both.} \rev{Section~3.2 takes up what this means for the channel separation.}

\subsection{Simulation details}

\rev{All results in this work were computed with the Tidy3D finite-difference time-domain solver.\cite{tidy3d} The normal-incidence spectra and the oblique-incidence sweeps come from a single model and a single mesh specification, so the resonance positions quoted in one place and used in another are the same numbers.}

\rev{The unit cell is periodic in $x$ and $y$ and terminated with perfectly matched layers along $z$.} \rev{Oblique incidence uses Bloch boundary conditions derived from the source, with the in-plane wavevector $k_x=(2\pi/\lambda)\sin\theta$.} \rev{The angle $\theta$ is stepped from $0^\circ$ to $12^\circ$ in $1^\circ$ increments, giving thirteen samples of the transfer function, with additional runs at $-6^\circ$ and $-12^\circ$ used to check the mirror symmetry of the result.} \rev{The base mesh is set by a target of 40 steps per wavelength, and the 60~nm gap is refined further to 4.3~nm across its width so that the anisotropic layer is resolved by about fourteen cells.} \rev{Transmission is recorded on a diffraction monitor placed in the substrate, which returns the complex zeroth-order amplitude, so the amplitude and the phase of the transfer function come from the same quantity in the same calculation.}

\rev{Mesh convergence was checked at 30, 40 and 50 steps per wavelength (Table~S1).} \rev{Between the last two settings the TE resonance moves by 0.33~nm and its quality factor by 2.8\%, and the TM resonance moves by 0.26~nm and its quality factor by 0.2\%.} \rev{Values quoted below are from the 40-step setting, and the residual numerical uncertainty is taken as $\pm0.3$~nm in wavelength and $\pm3\%$ in $Q$.} \rev{The geometric tolerance runs of Section~6.4 were carried out separately on a coarser grid, 22 steps per wavelength.} \rev{No transmittance value above unity occurs at any angle in the sweep.}

\rev{As a check on the solver, running the same geometry with the calculated constants of Lajaunie \textit{et al.}\cite{lajaunie2013} and with the stacking axis placed in the plane of the flake gives resonances at 886.9 and 913.5~nm, against 883.9 and 923.2~nm from an independent finite-difference calculation of the same configuration.} \rev{Agreement at this level is what the difference in numerical method would lead one to expect.}

\begin{figure}[tb]
\centering
\includegraphics[width=\columnwidth]{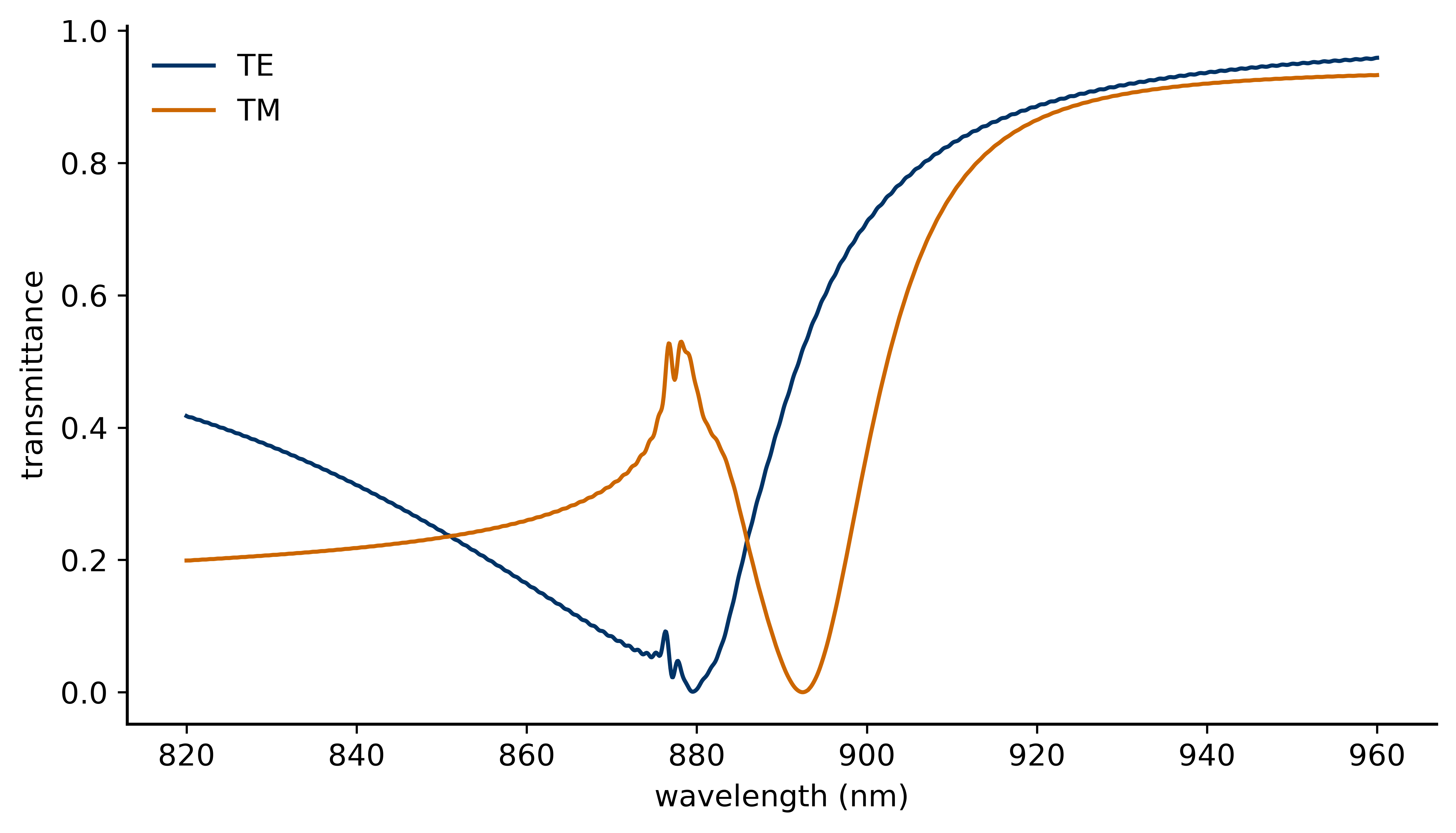}
\caption{\rev{Simulated transmittance at normal incidence for the two polarization channels, computed with the measured tensor and the exfoliated axis assignment.}}
\label{fig:spectra}
\end{figure}

\begin{table}[tb]
\small
  \caption{\ Fano fit parameters at normal incidence. $\lambda_0$: resonance wavelength. $Q$: quality factor. $q$: Fano asymmetry parameter. \rev{Values are quoted at 40 steps per wavelength; the numerical uncertainty from the convergence study of Table~S1 is $\pm0.3$~nm in $\lambda_0$ and $\pm3\%$ in $Q$}}
  \label{tbl:resonances}
  \begin{tabular*}{\columnwidth}{@{\extracolsep{\fill}}lll}
    \hline
    Parameter & TE channel & TM channel \\
    \hline
    Resonance wavelength $\lambda_0$ (nm) & 885.5 & 893.8 \\
    Quality factor $Q$                    & 50.5  & 44.8  \\
    Linewidth (nm)                        & 17.5  & 20.0  \\
    Fano asymmetry $q$                    & 0.74  & 0.14  \\
    Fano fit $R^2$                        & 0.986 & 0.999 \\
    Operating wavelength (nm)             & 880.8 & 893.8 \\
    Cutoff $\kappa$ ($\mu$m$^{-1}$)       & 0.436 & 1.165 \\
    Passband transmittance                & 0.70  & 0.15  \\
    \hline
  \end{tabular*}
\end{table}

\section{Fano resonance characterization}

\subsection{Normal-incidence spectra}

\rev{Fig.~\ref{fig:spectra} shows the simulated transmittance of both channels at normal incidence.} \rev{Each spectrum was fitted to the Fano lineshape\cite{limonov2017,fan2003}}\begin{equation}
  T(\lambda) = T_\mathrm{bg}\,
    \frac{\bigl[q + (\lambda-\lambda_0)/\Gamma\bigr]^2}
         {1 + \bigl[(\lambda-\lambda_0)/\Gamma\bigr]^2},
  \label{eq:fano}
\end{equation}
where $\lambda_0$ is the resonance wavelength, $\Gamma$ the half-linewidth with $Q=\lambda_0/2\Gamma$, $q$ the asymmetry parameter and $T_\mathrm{bg}$ the background. \rev{Fitted parameters are given in Table~\ref{tbl:resonances}.}

\rev{The TE resonance is at 885.5~nm with $Q=50.5$, and the TM resonance at 893.8~nm with $Q=44.8$, a separation of 8.3~nm.} \rev{The fits give $R^2$ of 0.986 and 0.999.}

\rev{We describe both features as Fano resonances rather than as quasi-BICs.} \rev{Two observations support the weaker term.} \rev{A diagonal permittivity tensor aligned with the structural axes preserves both mirror planes of the nanobar pair, so filling the gap does not break a protecting symmetry, and the mechanism that produces a symmetry-protected quasi-BIC is therefore absent.} \rev{The quality factors, 50.5 and 44.8, are also well below the values that motivate the quasi-BIC description.} \rev{The control series of Section~3.2 shows that an isotropic filler lowers the quality factor in the same way.} \rev{What the gap material does is load the resonator, and the lineshape that results is described by eqn~\eqref{eq:fano} with a finite asymmetry parameter in both channels.}

\rev{A second, narrower feature is present in the TE spectrum near 877~nm, on the short-wavelength flank of the main dip.} \rev{It is not used in what follows; the transfer functions of Section~4 are evaluated on the broad resonance.}

\rev{Including the measured extinction coefficients changes the TE quality factor by less than the fit resolution and the TM quality factor from 44.8 to 44.7.} \rev{Treating the difference as an absorptive contribution gives $Q_\mathrm{abs}\approx2\times10^{4}$ for the TM channel, more than two orders of magnitude above the radiative value, so both resonances are radiation limited over this band.}

\subsection{Contribution of the gap material to the channel separation}

\rev{An axis-aligned diagonal tensor preserves both mirror planes of the nanobar pair and cannot break the structural symmetry of the cell.} \rev{What it changes is the permittivity seen by each of the two mode families, which are already non-degenerate because $W\neq L$.} \rev{The two contributions are separated below by calculation.}

\rev{The same geometry was run with a series of gap fillings: an empty gap, two isotropic fillers, a uniaxial filler with the in-plane components averaged and the out-of-plane component retained, and the full measured tensor.} \rev{The results are given in Table~S2.} \rev{Replacing the empty gap with an isotropic filler of index comparable to the mean of the tensor gives a channel separation of 17.0~nm, larger than the 8.3~nm obtained with the measured tensor.} \rev{Retaining only the out-of-plane contrast gives 9.2~nm.} \rev{Adding the in-plane birefringence to that gives 8.3~nm.}

Three things follow. \rev{The empty cell already separates the two channels by 14.4~nm, so the shape anisotropy of the bar pair, and not the gap material, sets the scale of the effect.} \rev{Filling the gap then moves the separation in either direction depending on the index of the filler, to 5.8~nm at $n=1.46$ and to 17.0~nm at $n=2.230$.} \rev{Relative to the isotropic filler of matched index, the anisotropy of the crystal reduces the separation from 17.0 to 8.3~nm, of which the out-of-plane contrast accounts for 7.8~nm and the in-plane birefringence for 0.9~nm.} \rev{The in-plane birefringence is therefore the smallest of these terms, and an isotropic filler does not leave the two channels degenerate.}

\rev{The empty-gap entry in Table~S2 is reported for completeness only.} \rev{Its fitted TE quality factor exceeds $10^4$, which a time-domain calculation of the length used here cannot resolve, so that value should not be read quantitatively.}

\begin{figure*}[t]
\centering
\includegraphics[width=0.92\textwidth]{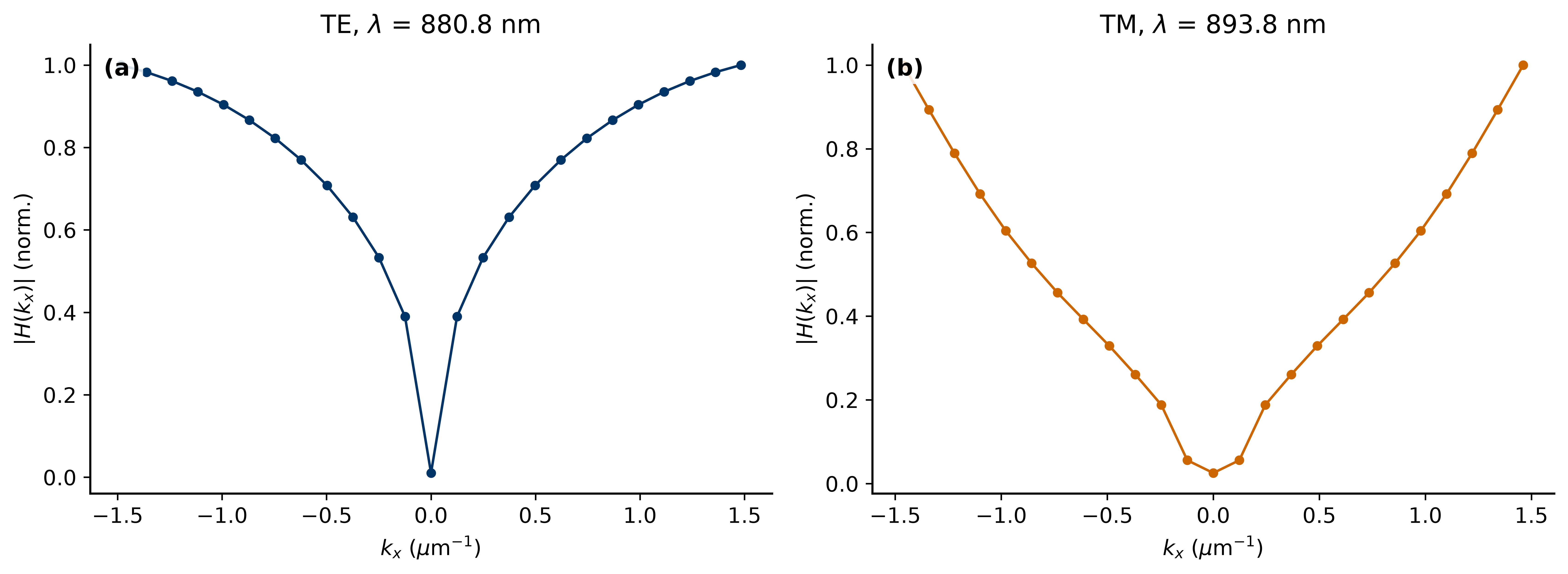}
\caption{\rev{Transfer function amplitude $|H(k_x)|$, normalised to its passband maximum. (a) TE channel at 880.8~nm. (b) TM channel at 893.8~nm. Both channels show a single minimum at $k_x=0$.}}
\label{fig:TF_amp}
\end{figure*}

\section{Spatial transfer functions}

\subsection{Extraction}

\rev{The diffraction monitor returns the complex zeroth-order transmission amplitude $t(k_x,\lambda)$.} \rev{The transfer function is taken as}\begin{equation}
  H(k_x) = \frac{t(k_x,\lambda_\mathrm{op})}{\max_{k_x}|t(k_x,\lambda_\mathrm{op})|},
  \label{eq:H}
\end{equation}
\rev{normalised to its own passband maximum at the operating wavelength.} \rev{A normalisation that instead divides the transmitted power by the transmission at a different wavelength uses a reference that is itself a function of angle, which distorts the extracted shape.} \rev{Absolute amplitudes are reported separately in Section~6.4 so that the efficiency remains visible.}

\rev{The operating wavelength is the minimum of $|t|$ at normal incidence on the 0.1~nm wavelength grid, giving 880.8~nm for TE and 893.8~nm for TM.} \rev{These are read from the computed spectrum and not from the Fano fit, so they need not coincide with the fitted $\lambda_0$ of Table~\ref{tbl:resonances}.}

\rev{Because the cell has a mirror plane normal to $x$, $t(+k_x)=t(-k_x)$, and $H(k_x)$ is an even function of $k_x$.} \rev{The computed amplitudes satisfy this to better than one part in $10^{10}$, and the runs at negative angle reproduce their positive counterparts to the same level.}

\subsection{TE channel}

\rev{The TE transfer function [Fig.~\ref{fig:TF_amp}(a)] has a single minimum, at $k_x=0$, where $|H|$ falls to $1.0\times10^{-2}$, and rises monotonically to unity at the edge of the swept range.} \rev{A single-null high-pass form,}\begin{equation}
  |H(k_x)| = \frac{|k_x|}{\sqrt{k_x^2+\kappa^2}},
  \label{eq:hp}
\end{equation}
\rev{fits the data with $\kappa_\mathrm{TE}=0.436~\mu\mathrm{m}^{-1}$ at $R^2=0.939$.} \rev{The cutoff is a fitted parameter of eqn~\eqref{eq:hp} over the full swept range rather than a directly resolved feature: $\kappa_\mathrm{TE}$ corresponds to $\theta=3.5^\circ$, so four of the swept angles, $0^\circ$ to $3^\circ$, lie below it.} \rev{Power-law forms in $|k_x|$ and $k_x^2$ do not describe this channel; both give negative $R^2$ over the same range (Table~S5).}

\rev{No second minimum appears anywhere in the swept range.}

\subsection{TM channel}

\rev{The TM transfer function [Fig.~\ref{fig:TF_amp}(b)] also has a single minimum at $k_x=0$, with $|H|=2.5\times10^{-2}$ there, but its cutoff is broader.} \rev{Eqn~\eqref{eq:hp} gives $\kappa_\mathrm{TM}=1.165~\mu\mathrm{m}^{-1}$ at $R^2=0.897$.} \rev{Because that cutoff is comparable to the largest momentum reached in the sweep, the response over the accessible range is close to linear: a fit of $|H|\propto|k_x|$ for $|k_x|\leq1.0~\mu\mathrm{m}^{-1}$ gives $R^2=0.986$, against 0.572 for a quadratic form.}

\rev{The two channels therefore carry out the same kind of operation with different cutoffs, the sharper cutoff belonging to the narrower resonance.} \rev{A transfer function proportional to $|k_x|$ is the kernel used for edge enhancement.}

\begin{figure*}[t]
\centering
\includegraphics[width=0.92\textwidth]{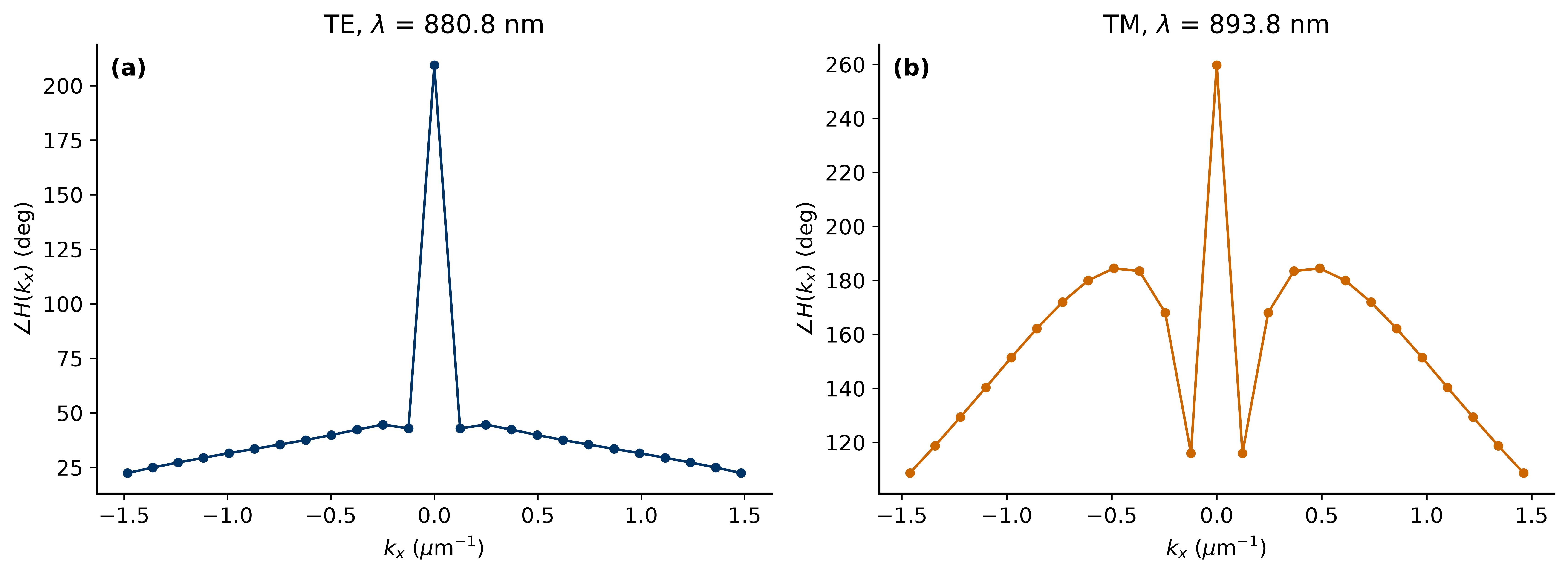}
\caption{\rev{Phase of the transfer function, obtained from the same complex amplitude as Fig.~\ref{fig:TF_amp}. (a) TE channel. (b) TM channel. The phase is even in $k_x$ in both channels.}}
\label{fig:phase}
\end{figure*}

\subsection{Phase}

The phase of $H(k_x)$ is shown in Fig.~\ref{fig:phase}. \rev{It is even in $k_x$ in both channels, as the mirror symmetry of the cell requires, and the computed values satisfy this to one part in $10^{10}$.}

\rev{This bears on how the TM operation should be described.} \rev{First-order spatial differentiation requires a transfer function proportional to $ik_x$, which is odd in $k_x$.} \rev{An even transfer function cannot take that form, so a metasurface with a mirror plane normal to $x$, illuminated at normal incidence, cannot implement $\partial/\partial x$.} \rev{The distinction lies in the sign structure of the phase rather than in its magnitude: differentiation requires $+90^\circ$ for $k_x>0$ and $-90^\circ$ for $k_x<0$, whereas a phase held at $+90^\circ$ on both sides, combined with a linear amplitude, gives $H\propto i|k_x|$, which is even and is not $\partial/\partial x$.} \rev{The amplitude behaviour of a first-order differentiator is reproduced here, but the phase is even, and the channel is better described as an edge-enhancement kernel with $|H|\propto|k_x|$.} \rev{The distinction affects the sign of the output rather than its magnitude, and the imaging results of Section~5 use the modulus.}

\rev{The phase plotted at $k_x=0$ carries no information in either channel, since $|H|$ there is $2.5\times10^{-2}$ or below and the argument of a near-zero amplitude is not well conditioned.}

\begin{figure*}[t]
\centering
\includegraphics[width=0.92\textwidth]{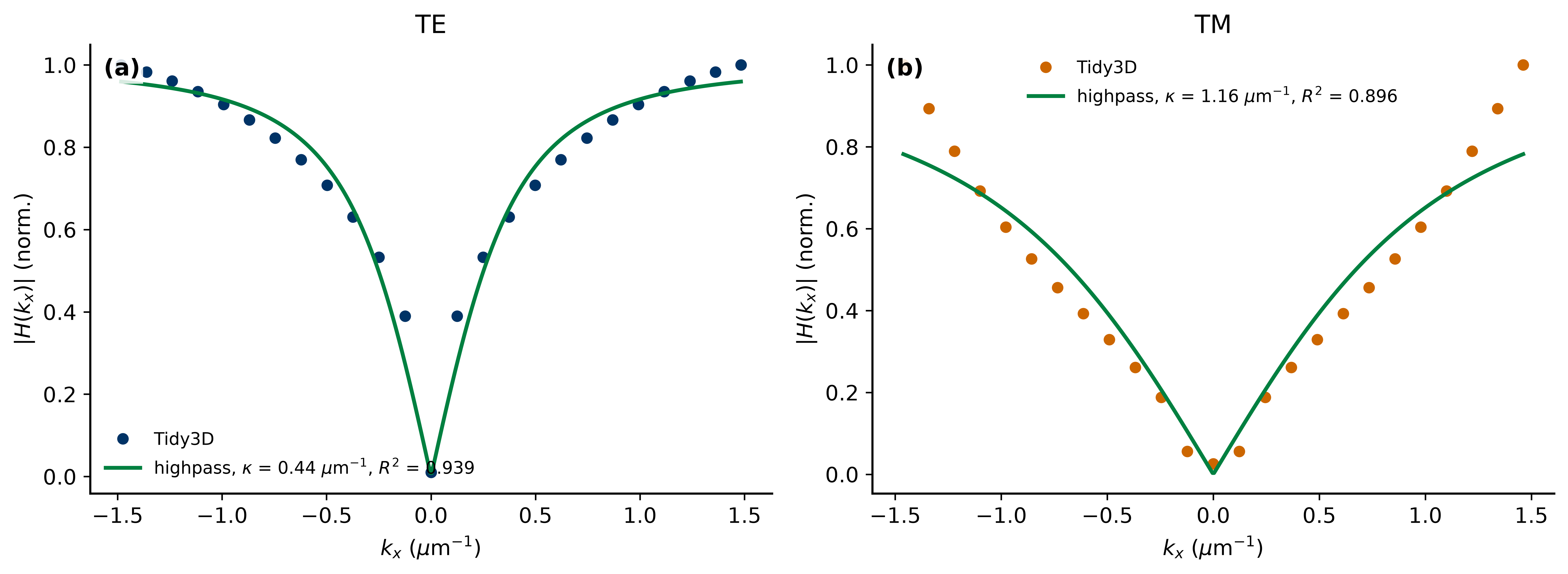}
\caption{\rev{Computed transfer function amplitude with the fitted single-null high-pass form of eqn~\eqref{eq:hp}. (a) TE channel, $\kappa=0.436~\mu$m$^{-1}$. (b) TM channel, $\kappa=1.165~\mu$m$^{-1}$. Fit statistics for all candidate models are given in Table~S5.}}
\label{fig:opid}
\end{figure*}

\subsection{Operation identification}

\rev{Fig.~\ref{fig:opid} compares both computed transfer functions with the fitted forms.} \rev{Fit parameters and coefficients of determination for all three candidate models in both channels are collected in Table~S5.}

\section{Image processing in a simulated 4$f$ arrangement}

Both operations were applied to a USAF~1951 resolution chart\cite{wiki_usaf1951} in a simulated 4$f$ arrangement.\cite{goodman2005} The input is placed at the front focal plane, the metasurface at the Fourier plane where it applies $H(k_x)$ as a multiplicative filter, and a second lens performs the inverse transform. \rev{The computed $H(k_x)$ of Section~4 is interpolated onto the spatial-frequency axis of the image and applied along $x$.} \rev{The object is sampled at the diffraction limit of the swept numerical aperture, $\lambda/2\mathrm{NA}=2.12~\mu$m per pixel, so that the spatial-frequency content of the $512\times512$ image lies inside the momentum range over which $H$ was computed; the corresponding field of view is 1.09~mm.} \rev{Sampling the object more finely than this would place image spatial frequencies outside the computed range, where the transfer function is not defined.} \rev{The filters used here are the simulated transfer functions themselves, not analytic models fitted to them.} \rev{Section~S9 of the Supplementary Information gives the sampling condition in full.}

\begin{figure*}[t]
\centering
\includegraphics[width=0.88\textwidth]{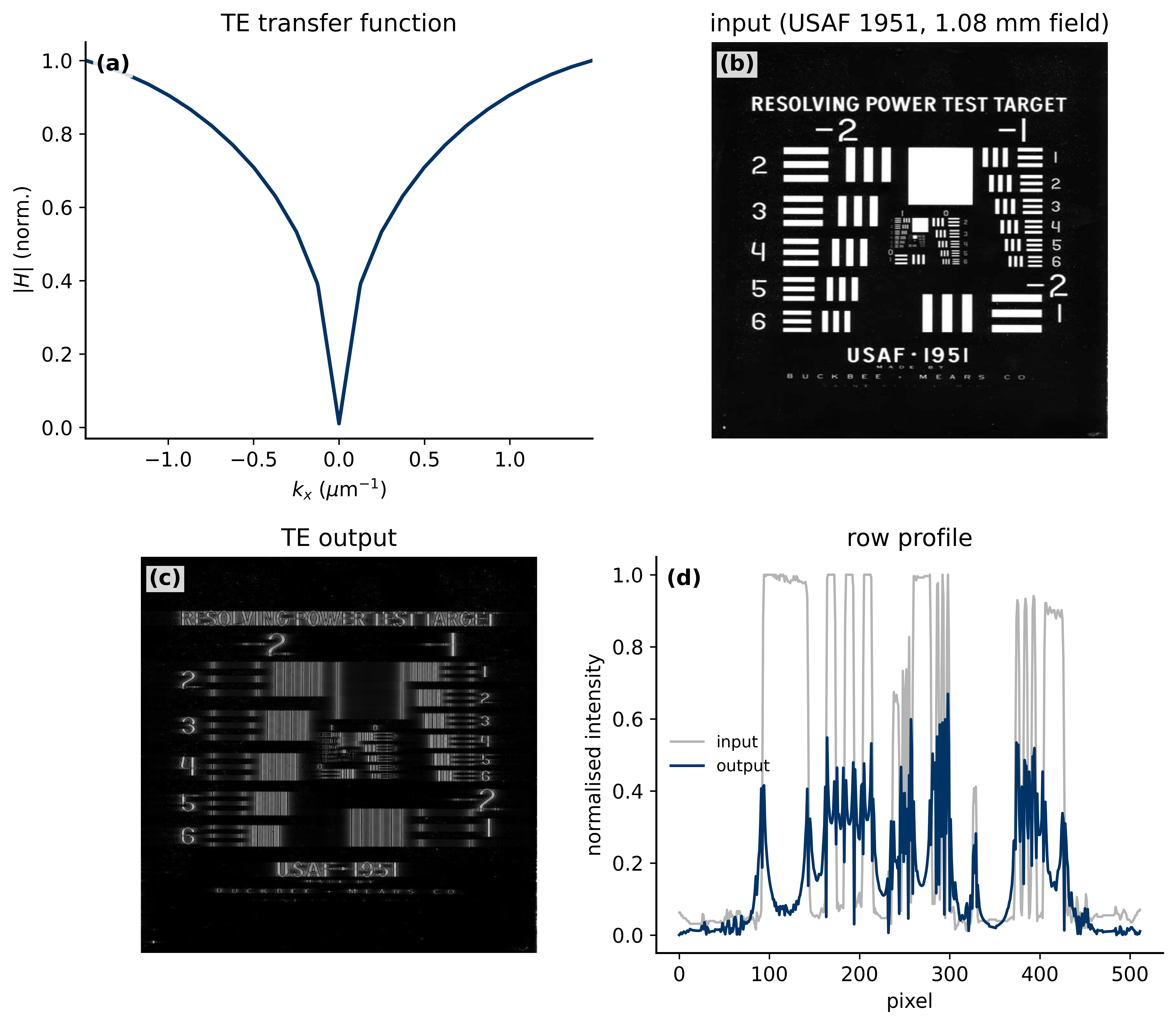}
\caption{Simulated 4$f$ image processing on a USAF~1951 resolution chart\cite{wiki_usaf1951} for the TE channel. \rev{(a) Transfer function applied in the Fourier domain. (b) Input image. (c) Output. (d) Row intensity profile through input and output.}}
\label{fig:imgTE}
\end{figure*}

\subsection{TE channel}

\rev{The TE filter suppresses the low spatial frequencies and passes the remainder with an amplitude that reaches unity at the edge of the computed range.} \rev{In the output [Fig.~\ref{fig:imgTE}] the bar boundaries of the chart appear as bright features against a dark background, and a row profile through the output shows a peak at each boundary.} \rev{The residual background visible in the output follows from the finite depth of the null, $|H(0)|=1.0\times10^{-2}$ rather than zero.}

\begin{figure*}[t]
\centering
\includegraphics[width=0.88\textwidth]{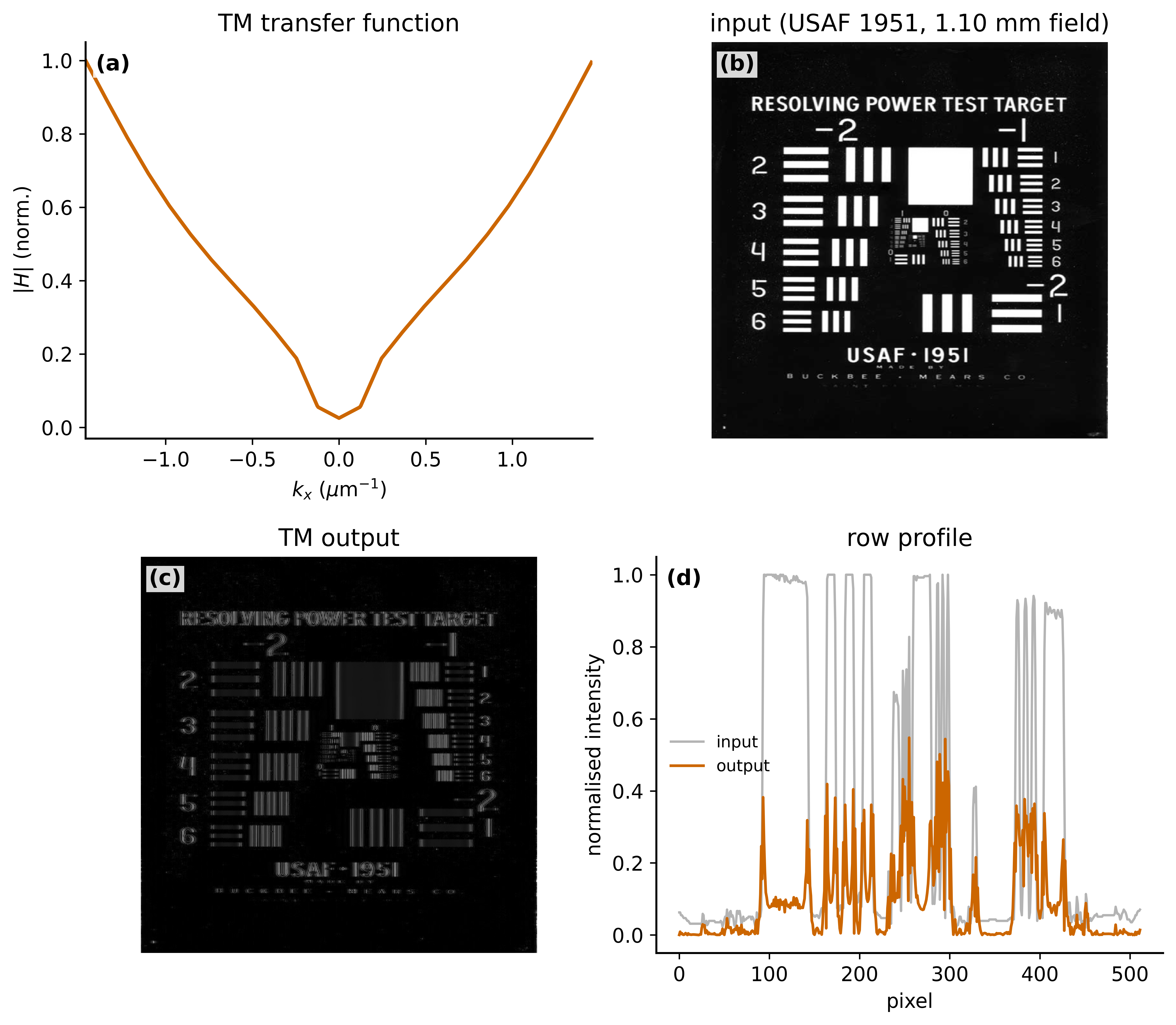}
\caption{Simulated 4$f$ image processing on a USAF~1951 resolution chart\cite{wiki_usaf1951} for the TM channel. \rev{(a) Transfer function applied in the Fourier domain. (b) Input image. (c) Output. (d) Row intensity profile through input and output.}}
\label{fig:imgTM}
\end{figure*}

\subsection{TM channel}

\rev{The TM filter, with $|H|\propto|k_x|$ over the accessible range, produces the output shown in Fig.~\ref{fig:imgTM}.} \rev{Transitions along $x$ appear as bright features and the row profile shows a peak at each edge.} Because the transfer function rises from zero rather than approaching unity in a passband, the overall output intensity is lower than in the TE case. \rev{This follows from the shape of the kernel, and it sets the optical budget for any implementation; Section~6.4 gives the corresponding numbers.}

\section{Discussion}

\subsection{Origin of the two operations}

\rev{Both channels have a transmission null at normal incidence and both act as spatial high-pass filters.} \rev{What separates them is the cutoff spatial frequency, $0.436~\mu\mathrm{m}^{-1}$ for TE against $1.165~\mu\mathrm{m}^{-1}$ for TM.} \rev{The broader cutoff belongs to the channel with the lower quality factor, 44.8 against 50.5, which is the expected ordering if the momentum-space width of the response is set by the linewidth of the resonance.} \rev{The two ratios are not equal, 2.7 against 1.1, so the ordering is reported as a trend and not as a scaling law.}

\rev{The practical difference between the channels follows from where the cutoff sits relative to the momentum range of interest.} \rev{The TE cutoff lies well inside that range, so the TE channel presents a sharp edge and a flat passband.} \rev{The TM cutoff lies near the top of it, so the TM channel remains in the part of the curve that is close to linear in $|k_x|$.}

\subsection{Role of the gap material}

\rev{Section~3.2 separates the contributions to the channel separation.} \rev{The shape anisotropy of the bar pair sets its scale, the out-of-plane index contrast of the crystal accounts for most of the anisotropic contribution, and the in-plane birefringence for about 0.9~nm.}

\rev{This bears on the choice of material.} \rev{Crystals with a larger in-plane anisotropy than $\alpha$-MoO$_3$ exist,\cite{caldwell2019,zheng2019} but in this geometry the in-plane birefringence is not the parameter that sets the channel separation, so a larger value of it would not by itself improve the device.} \rev{What the present arrangement uses is a low out-of-plane index in a high-index slot, together with a measured dielectric tensor in the operating band.} \rev{A geometry with four-fold rotational symmetry would give the in-plane birefringence the role assigned to it here, because the two channels would then be degenerate without it.} \rev{That is a direction for further work rather than a description of the structure studied here.}

\subsection{Channel isolation and polarization crosstalk}

\rev{At $0^\circ$ and $90^\circ$ input polarization the two channels are separated by the mirror symmetry of the cell.} If the crystal axes are rotated in the plane by an angle $\phi$ relative to the bar symmetry planes, an off-diagonal component $\varepsilon_{xy}$ appears and couples the two gap fields.

\rev{Table~S3 gives the computed leakage into the orthogonal polarization as a function of $\phi$, evaluated in the passband.} \rev{At $\phi=0$ the leaked power fraction is $2.1\times10^{-7}$, which is the numerical floor and is consistent with the exact zero that symmetry requires.} \rev{It rises to $5.5\times10^{-6}$ at $3^\circ$, $5.7\times10^{-5}$ at $10^\circ$ and $5.8\times10^{-4}$ at $45^\circ$.} \rev{These values lie well below the $\tfrac{1}{4}\sin^2(2\phi)$ estimate, so that estimate is conservative.} \rev{Maintaining $|\phi|<3^\circ$ keeps the leakage below $-50$~dB.} \rev{The in-plane orientation of the crystal axes would therefore have to be determined before placement, for example by polarimetric Raman mapping.}

\rev{Leakage evaluated at the operating wavelength instead of in the passband is larger, because the co-polarized amplitude there is small by construction.} \rev{Both sets of values are given in Table~S3.}

\subsection{Practical benchmarks}

\rev{The momentum range covered by the sweep is $|k_x|\leq1.48~\mu\mathrm{m}^{-1}$, corresponding to a numerical aperture of 0.208 at the TE operating wavelength.} \rev{In the passband the TE channel transmits a power fraction of 0.70 and the TM channel 0.15.} \rev{The lower TM figure follows from the shape of the kernel rather than from absorption, which Section~3.1 shows to be negligible here.} \rev{At the null the residual amplitudes are $1.0\times10^{-2}$ and $2.5\times10^{-2}$, which bound the contrast available from each channel.}

\rev{Table~S4 gives the sensitivity of both resonances to the geometric parameters.} \rev{A $\pm5$~nm change in the gap moves the TE resonance by about 2.5~nm and the TM resonance by less than 1~nm.} \rev{A 10~nm increase in bar width moves the TE resonance by 9.4~nm, so bar width is the tighter of the two lithographic tolerances.} \rev{Opening the gap to 100~nm shifts both resonances to longer wavelength and reduces the TE quality factor.} \rev{By the ordering of Section~6.1 that is the direction to move in if a broader cutoff is wanted.} \rev{Increasing the period to 650~nm raises both quality factors substantially; the TE value in that row exceeds the range this calculation can resolve and is quoted only as a direction.} \rev{These runs were performed on the coarser grid of 22 steps per wavelength noted in Section~2.3, so they should be read as shifts relative to their own nominal row rather than as absolute values.}

\begin{table*}[t]
\small
  \caption{\ Comparison with selected dual-channel analog computing demonstrations. HP: high-pass filter; D$_1$: first-order differentiator; D$_2$: second-order differentiator; Int: integrator; \rev{EE: edge-enhancement kernel with $|H|\propto|k_x|$. The present work is a simulation study; the other entries report experimental demonstrations}}
  \label{tbl:compare}
  \begin{tabular*}{\textwidth}{@{\extracolsep{\fill}}llll}
    \hline
    Reference & Operations & Channel control & Mechanism \\
    \hline
    Kwon \textit{et al.}\cite{kwon2018}       & D$_1$, D$_2$          & Polarization-free & Nonlocal, Fano \\
    Kwon \textit{et al.}\cite{kwon2020}       & Isotropic D$_1$/D$_2$ & Pol.-insensitive  & Nonlocal, Fano \\
    Zhou \textit{et al.}\cite{zhou2020}       & D$_1$, D$_2$          & Wavelength        & Geometric, resonant \\
    Zhou \textit{et al.}\cite{zhou2025qbic}   & Dual-pol.\ D$_2$      & Both pol.         & Geometric quasi-BIC \\
    Cotrufo \textit{et al.}\cite{cotrufo2023} & Dual-pol.\ D$_2$      & Polarization      & Dispersion engineering \\
    Bi \textit{et al.}\cite{bi2025}           & D$_1$ + Int           & Polarization      & Structural encoding \\
    \rev{This work (simulation)}              & \rev{HP at two cutoffs, EE} & \rev{Pol.\ angle} & \rev{Shape anisotropy, anisotropic gap fill} \\
    \hline
  \end{tabular*}
\end{table*}

\subsection{Comparison with related work}

Table~\ref{tbl:compare} places this work among selected multi-channel analog computing demonstrations.\cite{kwon2018,kwon2020,zhou2020,zhou2025qbic,cotrufo2023,bi2025} In most of those designs a separate structural element or a dispersion-engineered unit cell is used for each channel, and routing is by wavelength or by the orientation of the structural element. Here both channels share one unit cell and routing is by input polarization angle. \rev{The two operations available are high-pass filtering at two different cutoffs rather than two distinct mathematical operators, which is a narrower capability than a differentiator pair.} \rev{The comparison is between a simulation study and experimental demonstrations, and is offered only to place the operations and the routing mechanism side by side.}

\subsection{Requirements for an experimental realisation}

\rev{$\alpha$-MoO$_3$ flakes of 20 to 200~nm thickness can be prepared by mechanical exfoliation or chemical vapour deposition.\cite{caldwell2019} Filling a 60~nm gap requires either exfoliating a flake onto a pre-patterned substrate and placing it deterministically over the gap, or growing $\alpha$-MoO$_3$ by chemical vapour deposition after the bars are etched.} \rev{The two routes trade against each other.} \rev{Transfer allows the crystal axes to be checked and set before placement, which is what the $3^\circ$ tolerance of Section~6.3 requires, but a 60~nm high-aspect-ratio trench is not filled reliably by a transferred flake.} \rev{Growth after etching fills the trench, but on an amorphous surface it does not by itself fix the in-plane orientation of the crystal, so it would have to be combined with selection of correctly oriented grains or with an epitaxial substrate.} \rev{Neither route is straightforward at this gap width.}

\rev{The tolerances of Table~S4 apply to the gap width, the bar width and the period, and both resonances move by several nanometres for changes of a few nanometres in those dimensions.} \rev{Variation in how completely the gap is filled was not modelled, so its effect is not known from the present calculations.} \rev{The crystal-axis alignment tolerance of Section~6.3 is a separate requirement that the anisotropy imposes.}

\section*{Conclusions}

\rev{A symmetric TiO$_2$ nanobar-pair metasurface with a 60~nm $\alpha$-MoO$_3$ gap fill supports two Fano resonances, at 885.5~nm with $Q=50.5$ and at 893.8~nm with $Q=44.8$, in a single unmodified unit cell.} \rev{Both channels act as spatial high-pass filters and differ in their cutoff spatial frequency, $0.436$ and $1.165~\mu\mathrm{m}^{-1}$.} \rev{The broader channel remains close to linear in $|k_x|$ over the momentum range accessible here, with $R^2=0.986$, which is the regime used for edge enhancement.} \rev{Channel selection is by input polarization angle, and the results support the use of a single symmetric cell for two spatial filtering operations of different bandwidth.}

\rev{The shape anisotropy of the bar pair sets the scale of the channel separation: the empty cell already separates the two channels by 14.4~nm.} \rev{Relative to an isotropic filler of matched index the crystal reduces the separation from 17.0 to 8.3~nm, with the out-of-plane contrast accounting for 7.8~nm of that and the in-plane birefringence for 0.9~nm, so the biaxial character of $\alpha$-MoO$_3$ is not the dominant mechanism in this geometry.} \rev{A four-fold symmetric cell would give the in-plane birefringence a larger role.}

\rev{Two difficulties stand in the way of an experimental realisation.} \rev{Filling a 60~nm high-aspect-ratio gap and fixing the in-plane crystal orientation pull towards different preparation routes, as Section~6.6 sets out.} \rev{The TM channel transmits 0.15 of the incident power in its passband against 0.70 for the TE channel, which limits the optical budget available from that channel.} \rev{All results reported here are from simulation; experimental realisation and characterisation remain to be carried out.}

\section*{Author contributions}

S.\ Debnath: conceptualisation, methodology, software, formal analysis, investigation, visualisation, writing -- original draft, \rev{writing -- review and editing}. S.\ Saha: \rev{conceptualisation, methodology,} validation, supervision, \rev{resources,} writing -- review and editing. \rev{Both authors have read and approved the final manuscript.}

\section*{Conflicts of interest}

There are no conflicts to declare.

\section*{Data availability}

\rev{The data supporting this article are included in the manuscript and in the Supplementary Information. The unit-cell geometry, the material tensor, the mesh specification and the convergence, control, crosstalk, tolerance and fit data are given in Section~2 and in Tables~S1 to S5. The measured optical constants of $\alpha$-MoO$_3$ are those published by Toksumakov \textit{et al.}\cite{toksumakov2026}}

\section*{Acknowledgements}

The authors thank the Department of Electrical and Electronic Engineering, Bangladesh University of Engineering and Technology for computational resources. \rev{The authors thank the anonymous referees for their constructive comments.}

\balance

\bibliography{moo3_differentiator_ol}

@article{koshelev2018,
  title={Asymmetric metasurfaces with high-Q resonances governed by bound states in the continuum},
  author={Koshelev, Kirill and Lepeshov, Sergey and Liu, Mingkai and Bogdanov, Andrey and Kivshar, Yuri},
  journal={Physical Review Letters},
  volume={121},
  number={19},
  pages={193903},
  year={2018},
  publisher={APS},
  doi={10.1103/PhysRevLett.121.193903}
}

@article{hsu2016,
  title={Bound states in the continuum},
  author={Hsu, Chia Wei and Zhen, Bo and Stone, A Douglas and Joannopoulos, John D and Solja{\v{c}}i{\'c}, Marin},
  journal={Nature Reviews Materials},
  volume={1},
  number={9},
  pages={16048},
  year={2016},
  publisher={Nature Publishing Group}
}

@article{silva2014,
  title={Performing mathematical operations with metamaterials},
  author={Silva, Alexandre and Monticone, Francesco and Castaldi, Giuseppe and Galdi, Vincenzo and Al{\`u}, Andrea and Engheta, Nader},
  journal={Science},
  volume={343},
  number={6167},
  pages={160--163},
  year={2014},
  publisher={American Association for the Advancement of Science}
}

@article{zangeneh2021,
  title={Analogue computing with metamaterials},
  author={Zangeneh-Nejad, Farzad and Sounas, Dimitrios L and Al{\`u}, Andrea and Fleury, Romain},
  journal={Nature Reviews Materials},
  volume={6},
  number={3},
  pages={207--225},
  year={2021},
  publisher={Nature Publishing Group UK London}
}

@article{cordaro2019,
  title={High-index dielectric metasurfaces performing mathematical operations},
  author={Cordaro, Andrea and Kwon, Hoyeong and Sounas, Dimitrios and Koenderink, A Femius and Al{\`u}, Andrea and Polman, Albert},
  journal={Nano Letters},
  volume={19},
  number={12},
  pages={8418--8423},
  year={2019},
  publisher={ACS Publications}
}

@article{zhu2017,
  title={Plasmonic computing of spatial differentiation},
  author={Zhu, Tengfeng and Zhou, Yihan and Lou, Yijie and Ye, Hui and Qiu, Min and Ruan, Zhichao and Fan, Shanhui},
  journal={Nature Communications},
  volume={8},
  number={1},
  pages={15391},
  year={2017},
  publisher={Nature Publishing Group UK London}
}

@article{guo2018,
  title={Photonic crystal slab {L}aplace operator for image differentiation},
  author={Guo, Cheng and Xiao, Meng and Minkov, Momchil and Shi, Yu and Fan, Shanhui},
  journal={Optica},
  volume={5},
  number={3},
  pages={251--256},
  year={2018},
  publisher={Optical Society of America}
}

@article{zhou2020,
  title={Flat optics for image differentiation},
  author={Zhou, You and Zheng, Hanyu and Kravchenko, Ivan I and Valentine, Jason},
  journal={Nature Photonics},
  volume={14},
  number={5},
  pages={316--323},
  year={2020},
  publisher={Nature Publishing Group UK London}
}

@article{zong2023,
  title={Two-dimensional optical differentiator for broadband edge detection based on dielectric metasurface},
  author={Zong, Meixue and Liu, Yiqing and Lv, Jinwen and Zhang, Shubin and Xu, Zhengji},
  journal={Optics Letters},
  volume={48},
  number={7},
  pages={1902--1905},
  year={2023},
  publisher={Optica Publishing Group}
}

@article{cotrufo2023,
  title={Dispersion engineered metasurfaces for broadband, high-{NA}, high-efficiency, dual-polarization analog image processing},
  author={Cotrufo, Michele and Arora, Akshaj and Singh, Sahitya and Al{\`u}, Andrea},
  journal={Nature Communications},
  volume={14},
  number={1},
  pages={7078},
  year={2023},
  publisher={Nature Publishing Group UK London}
}

@article{deng2024,
  title={Broadband angular spectrum differentiation using dielectric metasurfaces},
  author={Deng, Ming and Cotrufo, Michele and Wang, Jian and Dong, Jianji and Ruan, Zhichao and Al{\`u}, Andrea and Chen, Lin},
  journal={Nature Communications},
  volume={15},
  number={1},
  pages={2237},
  year={2024},
  publisher={Nature Publishing Group UK London}
}

@article{limonov2017,
  title={{F}ano resonances in photonics},
  author={Limonov, Mikhail F and Rybin, Mikhail V and Poddubny, Alexander N and Kivshar, Yuri S},
  journal={Nature Photonics},
  volume={11},
  number={9},
  pages={543--554},
  year={2017},
  publisher={Nature Publishing Group UK London}
}

@article{swartz2024,
  title={Broadband and large-aperture metasurface edge encoders for incoherent infrared radiation},
  author={Swartz, Brandon T and Zheng, Hanyu and Forcherio, Gregory T and Valentine, Jason},
  journal={Science Advances},
  volume={10},
  number={6},
  pages={eadk0024},
  year={2024},
  publisher={American Association for the Advancement of Science}
}

@article{zhou2019,
  author={Zhou, J. and Qian, H. and Chen, C.-F. and Zhao, J. and Li, G. and Wu, Q. and Luo, H. and Wen, S. and Liu, Z.},
  title={Optical edge detection based on high-efficiency dielectric metasurface},
  journal={Proceedings of the National Academy of Sciences},
  volume={116},
  number={23},
  pages={11137--11140},
  year={2019},
  doi={10.1073/pnas.1820636116}
}

@article{yang2025,
  title={Permittivity-asymmetric q{BIC} metasurfaces for refractive index sensing},
  author={Yang, Xingye and Antonov, Alexander and Hu, Haiyang and Tittl, Andreas},
  journal={Nanophotonics},
  volume={14},
  number={27},
  pages={5311--5321},
  year={2025},
  publisher={De Gruyter}
}

@article{kischkat2012,
  title={Mid-infrared optical properties of thin films of aluminum oxide, titanium dioxide, silicon dioxide, aluminum nitride, and silicon nitride},
  author={Kischkat, Jan and Peters, Sven and Gruska, Bernd and Semtsiv, Mykhaylo and Chashnikova, Mikaela and Klinkm{\"u}ller, Matthias and Fedosenko, Oliana and Machulik, Stephan and Aleksandrova, Anna and Monastyrskyi, Gregorii and others},
  journal={Applied Optics},
  volume={51},
  number={28},
  pages={6789--6798},
  year={2012},
  publisher={Optical Society of America}
}

@article{horton2025accelerated,
  title={Accelerated data-driven materials science with the Materials Project},
  author={Horton, Matthew K and Huck, Patrick and Yang, Ruo Xi and Munro, Jason M and Dwaraknath, Shyam and Ganose, Alex M and Kingsbury, Ryan S and Wen, Mingjian and Shen, Jimmy X and Mathis, Tyler S and others},
  journal={Nature Materials},
  volume={24},
  number={10},
  pages={1522--1532},
  year={2025},
  publisher={Nature Publishing Group UK London}
}

@article{jain2013commentary,
  author = {Jain, Anubhav and Ong, Shyue Ping and Hautier, Geoffroy and Chen, Wei and Richards, William Davidson and Dacek, Stephen and Cholia, Shreyas and Gunter, Dan and Skinner, David and Ceder, Gerbrand and Persson, Kristin A.},
    title = {Commentary: The Materials Project: A materials genome approach to accelerating materials innovation},
    journal = {APL Materials},
    volume = {1},
    number = {1},
    pages = {011002},
    year = {2013},
    month = {07},issn = {2166-532X},
    doi = {10.1063/1.4812323},
    url = {https://doi.org/10.1063/1.4812323},
    eprint = {https://pubs.aip.org/aip/apm/article-pdf/doi/10.1063/1.4812323/13163869/011002_1_online.pdf}
}

@misc{wiki_usaf1951,
  author={M. Ryazanov},
  title={{USAF-1951 Resolution Test Chart}},
  year={2022},
  howpublished={\url{https://commons.wikimedia.org/wiki/File:USAF-1951.svg}},
  note={Wikimedia Commons, accessed May 2026}
}

@book{goodman2005,
  author    = {Goodman, Joseph W.},
  title     = {Introduction to Fourier Optics},
  edition   = {4th},
  year      = {2017},
  publisher = {W. H. Freeman},
  address   = {New York, NY},
  isbn      = {978-1319119164}
}

@article{fan2003,
  title={Temporal coupled-mode theory for the {F}ano resonance in optical resonators},
  author={Fan, Shanhui and Suh, Wonjoo and Joannopoulos, John D},
  journal={Journal of the Optical Society of America A},
  volume={20},
  number={3},
  pages={569--572},
  year={2003},
  publisher={Optical Society of America}
}

@article{caldwell2019,
  title={Photonics with hexagonal boron nitride},
  author={Caldwell, Joshua D and Aharonovich, Igor and Cassabois, Guillaume and Edgar, James H and Gil, Bernard and Basov, D N},
  journal={Nature Reviews Materials},
  volume={4},
  number={8},
  pages={552--567},
  year={2019},
  publisher={Nature Publishing Group UK London}
}

@article{gomis2017,
  title={Anisotropy-induced photonic bound states in the continuum},
  author={Gomis-Bresco, Jordi and Artigas, David and Torner, Lluis},
  journal={Nature Photonics},
  volume={11},
  number={4},
  pages={232--236},
  year={2017},
  publisher={Nature Publishing Group UK London}
}

@article{tittl2018,
  title={Imaging-based molecular barcoding with pixelated dielectric metasurfaces},
  author={Tittl, Andreas and Leitis, Aleksandrs and Liu, Mingkai and Yesilkoy, Filiz and Choi, Duk-Yong and Neshev, Dragomir N and Kivshar, Yuri S and Altug, Hatice},
  journal={Science},
  volume={360},
  number={6393},
  pages={1105--1109},
  year={2018},
  publisher={American Association for the Advancement of Science}
}

@article{he2022,
  title={Computing metasurfaces for all-optical image processing: a brief review},
  author={He, Shanshan and Wang, Ruisi and Luo, Hailu},
  journal={Nanophotonics},
  volume={11},
  number={6},
  pages={1083--1108},
  year={2022},
  publisher={De Gruyter}
}

@article{bi2025,
  title={Concurrent image differentiation and integration processings enabled by polarization-multiplexed metasurface},
  author={Bi, Xinyi and Wu, Xuanguang and Fan, Xinhao and Zhao, Chenyang and Wen, Dandan and Liu, Sheng and Gan, Xuetao and Zhao, Jianlin and Li, Peng},
  journal={Laser \& Photonics Reviews},
  volume={19},
  number={4},
  pages={2400718},
  year={2025},
  publisher={Wiley Online Library}
}

@article{kwon2018,
  title={Nonlocal metasurfaces for optical signal processing},
  author={Kwon, Hoyeong and Sounas, Dimitrios and Cordaro, Andrea and Polman, Albert and Al{\`u}, Andrea},
  journal={Physical review letters},
  volume={121},
  number={17},
  pages={173004},
  year={2018},
  publisher={APS}
}

@article{kwon2020,
  title={Dual-polarization analog 2D image processing with nonlocal metasurfaces},
  author={Kwon, Hoyeong and Cordaro, Andrea and Sounas, Dimitrios and Polman, Albert and Al{\`u}, Andrea},
  journal={Acs Photonics},
  volume={7},
  number={7},
  pages={1799--1805},
  year={2020},
  publisher={ACS Publications}
}

@article{kang2023,
  title={Applications of bound states in the continuum in photonics},
  author={Kang, Meng and Liu, Tao and Chan, Che Ting and Xiao, Meng},
  journal={Nature Reviews Physics},
  volume={5},
  number={11},
  pages={659--678},
  year={2023},
  publisher={Nature Publishing Group UK London}
}

@article{cui2026,
  title={Metasurfaces for edge detection and spatial differentiation in free space},
  author={Cui, Xingzhe and He, Shengjie and Li, Zhongjun and Chen, Tao and Zhang, Xu and Sun, Yuxiang and Wu, Yunkai and Guan, Jun and Hu, Jingtian},
  journal={Advanced Functional Materials},
  pages={e74788},
  year={2026},
  publisher={Wiley Online Library}
}

@article{zhou2025qbic,
  title={Dual-Polarized Broadband Laplace Differentiator via Quasi-Bound States in the Continuum Empowered by Nonlocal Metasurfaces},
  author={Zhou, Chen and Zhao, Ruizhe and Li, Peijin and Zhang, Yan and Chen, Yanjie and Geng, Guangzhou and Li, Junjie and Li, Xiaowei and Wang, Yongtian and Huang, Lingling},
  journal={Advanced Functional Materials},
  volume={35},
  number={37},
  pages={2426095},
  year={2025},
  publisher={Wiley Online Library}
}

@article{lajaunie2013,
  title={Strong anisotropic influence of local-field effects on the dielectric response of $\alpha$-MoO 3},
  author={Lajaunie, L and Boucher, F and Dessapt, R and Moreau, P},
  journal={Physical Review B},
  volume={88},
  number={11},
  pages={115141},
  year={2013}
}

@article{zheng2019,
  title={A mid-infrared biaxial hyperbolic van der Waals crystal},
  author={Zheng, Zebo and Xu, Ningsheng and Oscurato, Stefano L and Tamagnone, Michele and Sun, Fengsheng and Jiang, Yinzhu and Ke, Yanlin and Chen, Jianing and Huang, Wuchao and Wilson, William L and others},
  journal={Science advances},
  volume={5},
  number={5},
  pages={eaav8690},
  year={2019},
  publisher={American Association for the Advancement of Science}
}

@article{pan2021,
  title={Laplace metasurfaces for optical analog computing based on quasi-bound states in the continuum},
  author={Pan, Danping and Wan, Lei and Ouyang, Min and Zhang, Wei and Potapov, Alexander A and Liu, Weiping and Liang, Zixian and Feng, Tianhua and Li, Zhaohui},
  journal={Photonics Research},
  volume={9},
  number={9},
  pages={1758--1766},
  year={2021},
  publisher={Chinese Laser Press and Optical Society of America}
}

@article{liu2024,
  title={Edge detection imaging by quasi-bound states in the continuum},
  author={Liu, Tingting and Qiu, Jumin and Xu, Lei and Qin, Meibao and Wan, Lipeng and Yu, Tianbao and Liu, Qiegen and Huang, Lujun and Xiao, Shuyuan},
  journal={Nano Letters},
  volume={24},
  number={45},
  pages={14466--14474},
  year={2024},
  publisher={ACS Publications}
}

@article{miroshnichenko2010,
  title={Fano resonances in nanoscale structures},
  author={Miroshnichenko, Andrey E and Flach, Sergej and Kivshar, Yuri S},
  journal={Reviews of Modern Physics},
  volume={82},
  number={3},
  pages={2257--2298},
  year={2010},
  publisher={APS}
}

@article{chern2023,
  title={Bound states in the continuum in anisotropic photonic crystal slabs},
  author={Chern, Ruey-Lin and Chang, Jui-Chien and Yang, Hsueh-Chi},
  journal={Scientific Reports},
  volume={13},
  number={1},
  pages={14139},
  year={2023},
  publisher={Nature Publishing Group UK London}
}

@article{gupta2026,
  title={Anisotropy-induced Fano resonance and strong coupling in a two-dimensional black phosphorus-based infrared metasurface},
  author={Gupta, Vivek Kumar and Yadav, Narendra Kumar and Gupta, Prince and Pal, Satyendra Prakash and Singh, Vivek and Dayal, Govind},
  journal={Journal of the Optical Society of America B},
  volume={43},
  number={3},
  pages={581--587},
  year={2026},
  publisher={Optica Publishing Group}
}

@article{kendall2025,
  title={Dynamically reconfigurable 2D polarization-agnostic image edge-detection using nonvolatile phase-change metasurfaces},
  author={Kendall, Stuart and Ruiz de Galarreta, Carlota and Shields, Joe and Bertolotti, Jacopo and Yang, Guoce and Wang, Mengyun and Al{\`u}, Andrea and Bhaskaran, Harish and Wright, C David},
  journal={Optics Express},
  volume={33},
  number={4},
  pages={8971--8982},
  year={2025},
  publisher={Optica Publishing Group}
}

@article{yu2026,
  title={Double-phase metasurface operators for all-optical image processing},
  author={Yu, Linzhi and Singh, Haobijam J and Pietila, Jesse and Caglayan, Humeyra},
  journal={Light: Science \& Applications},
  volume={15},
  number={1},
  pages={119},
  year={2026},
  publisher={Nature Publishing Group UK London}
}

@misc{tidy3d,
  title        = {{Tidy3D}: {GPU}-accelerated finite-difference time-domain
                  electromagnetic solver},
  author       = {{Flexcompute Inc.}},
  year         = {2026},
  note         = {Version 2.12.0},
  howpublished = {\url{https://www.flexcompute.com/tidy3d/}}
}

@article{toksumakov2026,
  title={Unveiling $\alpha$-{MoO}$_3$ as a High-Index, Low-Loss Anisotropic Material:
         Bridging the Gap From Mid-Infrared to Ultraviolet Nanophotonics},
  author={Toksumakov, Adilet N. and Grudinin, Dmitriy V. and Ermolaev, Georgy A. and
          Tsapenko, Alexey P. and Minnekhanov, Anton A. and Kravtsov, Konstantin V. and
          Pak, Nikolay V. and Mazitov, Arslan B. and Vyshnevyy, Andrey A. and
          Slavich, Aleksandr S. and Kruglov, Ivan A. and Tselikov, Gleb I. and
          Arsenin, Aleksey V. and Novoselov, Kostya S. and Volkov, Valentyn S.},
  journal={Advanced Materials},
  volume={38},
  pages={e74228},
  year={2026},
  publisher={Wiley},
  doi={10.1002/adma.74228}
}
\bibliographystyle{rsc}

\end{document}